\documentclass{vgtc}                          

\usepackage[lined, ruled, vlined]{algorithm2e}
\usepackage{amsfonts}
\usepackage{graphicx}

\usepackage{wasysym}
\usepackage{textcomp}
\usepackage{times}

\usepackage{latexsym}
\usepackage{epsfig}
\usepackage{psboxit}
\usepackage{wrapfig}
\usepackage{subfigure}
\usepackage{amsfonts}
\usepackage{graphicx}

\usepackage{url}

\onlineid{118}

\vgtccategory{Research}

\vgtcinsertpkg

\title{A Force-Directed Method for Large Crossing Angle Graph Drawing}

\author{Peter Eades\thanks{e-mail: peter@it.usyd.edu.au}\\ %
        \scriptsize University of Sydney %
\and Weidong Huang\thanks{e-mail: whua5569@it.usyd.edu.au}\\ %
     \scriptsize  University of Sydney %
\and Seok-Hee Hong\thanks{e-mail: shhong@it.usyd.edu.au}\\ %
     \scriptsize University of Sydney}

\abstract{

Recent empirical research has indicated that human graph reading performance improves when crossing angles increase. However, \emph{crossing angle} has not been used as an aesthetic criterion for graph drawing algorithms so far. In this paper, we introduce a force-directed method that aims to construct graph drawings with large crossing angles. Experiments indicate that our method significantly increases crossing angles. Surprisingly, the experimental results further demonstrate that the resulting drawings produced by our method have fewer edge crossings, a shorter total edge length and more uniform edge lengths, compared to classical spring algorithms.

} 


\CCScatlist{
  \CCScat{G.2.2}{Discrete Mathematics}{Graph Theory}{ Graph Algorithms}}

\begin{document}


\firstsection{Introduction}

\maketitle

A great deal of real world data have a relational structure and can be modeled as graphs. Graphs are usually drawn as node-link diagrams so that humans can make sense of the underlying structure. The past two decades have seen a fast growing body of research dedicated to designing algorithms to construct aesthetically pleasing drawings of graphs~\cite{di}.  While judgement of the quality of a drawing is subjective, a number of aesthetic criteria are generally accepted, including the following:

\begin{itemize}
\item Small number of edge crossings
\item Even distribution of vertices
\item Uniform edge length
\item Small drawing area
\item Maximum angular resolution
\end{itemize}

\begin{figure}[t]
 \centering
 \includegraphics[width=3.2in]{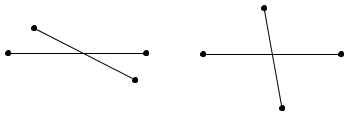}
 \caption{Theoretical and empirical research reveals that the crossing on the left is more confusing than that on the right.}
 \label{fig:eye:crossinangle}
\end{figure}

These criteria were originally proposed based on human intuition. However, some have been validated in user studies, mainly conducted by Purchase et al. (e.g.,~\cite{pur}), indicating that drawings satisfying such criteria yield some insights from the end user's point of view. For example, edge crossings were found to have the greatest impact on human graph reading performance. Recently, researchers have begun investigating theories from general psychology and adopting empirical methods in order to develop human-centered criteria for graph drawing~\cite{van,ware}. Among them, Ware et al.~\cite{ware} studied results from neurophysiology and suggested that edges that
cross at nearly 90 degrees are less likely to be confusing than those crossing at acute angles (see Figure~\ref{fig:eye:crossinangle}). The effect of crossing angle was subsequently observed in a qualitative eye tracking study by Huang~\cite{hu07}. This was validated quantitatively in a controlled experiment~\cite{hu08}. To be more specific, it was found that human graph reading performance improves when crossing angle increases.

While a great deal of attention has been focused on reducing the number of crossings, little research has been done on how to handle the remaining crossings. Didimo et al.~\cite{didimo} initiated a study of combinatorial questions related to drawing graphs with right angle crossings. Dunne and Shneiderman~\cite{dun} list crossing angle as a \lq\lq readability metric\rq\rq. However, in this previous research, no algorithm for producing graph drawings with large crossing angles has been proposed. In this paper, we introduce a force-directed method that aims to construct graph drawings with large crossing angles, called BIGCROSS. Experiments indicate that our method significantly increases crossing angles. Surprisingly, our experimental results further demonstrate that the resulting drawings produced by our method also have fewer edge crossings, a shorter total edge length and more uniform edge lengths, compared to a classical spring algorithm.

\section{The Classical Spring Algorithm}               \label{class}

\begin{figure}[h]
\centering
\includegraphics[width=3.4in]{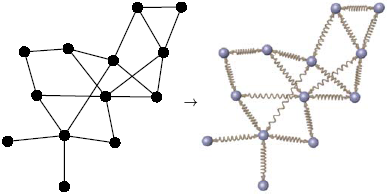}
\caption{The spring embedder model (taken from Brandes~\cite{brandes})} \label{fig:model}
\end{figure}

Force-directed graph drawing has been studied extensively (e.g.,~\cite{eades,fr}). In the \emph{ classical spring algorithm}, a graph is modeled as a physical system, in which vertices are replaced with steel rings, and edges are replaced with rings (see Figure~\ref{fig:model}). Springs pull connected vertices toward to each other when stretched, while they push vertices apart when compressed, following Hooke's law:
\begin{equation}
f_{s} = k_{s}(d-l).
\end{equation}
The repulsive force between all rings follows an inverse square law:
\begin{equation}
f_{r} = k_{r}/d^2.
\end{equation}
Here $k_{s}$ and $k_{r}$ are constants, $d$ is the Euclidean distance between vertices and $l$ is the natural length of the spring.

Starting with an arbitrary placement of vertices, the algorithm calculates the combined force on each vertex and moves the vertices accordingly. This process is repeated for a fixed number of times and the resulting pictures are usually aesthetically pleasing.

In the last two decades, the spring embedder model has been refined and extended in many different ways. Gansner and North~\cite{north} introduced two post-processing techniques to avoid vertex-vertex and vertex-edge overlaps. Brandes and Wagner~\cite{br} defined a random field model for drawing so-called \emph{train graphs} to avoid overlaps and small angles between edges resulting from straight-line edges. Lin and Yen~\cite{lin} introduced a repulsive force between adjacent edges to improve angular resolution. To visually separate clusters, Huang et al.~\cite{mlhuang} introduced dummy vertices representing individual clusters and repulsive forces that act between the dummy vertices. Recently, Hu and Koren~\cite{hu} proposed and implemented two post-processing algorithms to overcome warping effects resulted from the spring embedder model. Holten and van Wijk~\cite{holten} introduced a self-organizing approach in which edges are modeled as flexible springs that can attract each other for edge bundling.

More sophisticated techniques have been proposed for various purposes (e.g.,~\cite{hong,hu1,itoh,kobourov}). For excellent reviews, see~\cite{brandes,di}.

\section{The BIGCROSS Method} \label{sec:approach}

The beauty of force-directed graph drawing is that we can simply add up energy functions, each of which aims to maximize a specific aesthetic criterion. Our new ``BIGCROSS'' method, defined below, is an extension of the classical spring algorithm. We define a new force that increases the angle between crossing edges, and apply it in addition to the forces of the classical model.

\subsection{BIGCROSS Force Magnitude}

 The BIGCROSS algorithm extends to introduce an extra force called the \emph{cosine force} to increase crossing angle. The basic idea on how the cosine force works is as follows.

\begin{figure}[h]
\centering
\includegraphics[width=2.3in]{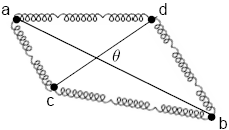}
\caption{The model of cosine force} \label{fig:cosforce1}
\end{figure}

As shown in Figure~\ref{fig:cosforce1}, if two edges $(a,b)$ and $(c,d)$ cross, then their endpoints $a$, $b$, $c$ and $d$ will be connected by special springs. These springs are special in that they work together and apply a non-Hooke's-law force on each of the endpoints, in such a way that when the crossing angle increases, energy decreases. To be more specific, each spring exerts forces on its connected vertices according to the crossing angle it faces. If the angle is acute, then the spring push the vertices apart. If the angle is obtuse, then the vertices are pulled toward each other. If it is a right angle, no force is applied. The magnitude of the force is $k_{cos}\cos\theta$, where $\theta$ ($\theta<90$\textdegree) is the crossing angle between the two edges, and $k_{cos}$ is a constant.

\subsection{BIGCROSS Force Direction}

During our investigation, we experimented with three different directions in which the cosine force may force crossings toward the right angle.

\begin{figure}[h]
\centering
\includegraphics[width=2.2in]{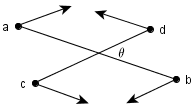}
\caption{The parallel cosine force} \label{fig:cosine1}
\end{figure}

The first one is that the force is applied in the direction of the other crossed edge. To be more specific, suppose that the position of vertex $v$ in an 2D Euclidean space is denoted by $p_v=(x_v,y_v)$, and the distance between vertices $u$ and $v$ is $d_{uv}$; then given two crossing edges $(a,b)$ and $(c,d)$ as shown in Figure~\ref{fig:cosine1}, the cosine force on vertex $a$, which we call \emph{parallel cosine force}, can be denoted as follows:

\begin{equation}
 (k_{cos}\cos\theta\frac{x_{d}-x_{c}}{d_{cd}}, k_{cos}\cos\theta\frac{y_{d}-y_{c}}{d_{cd}})
 \end{equation}

\begin{figure}[h]
\centering
\includegraphics[width=2.2in]{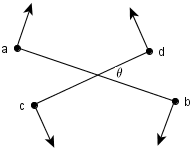}
\caption{The rotational cosine force} \label{fig:cosine2}
\end{figure}

In the second case, the force is applied in the direction orthogonal to the crossed edge. This is a kind of rotational force. Given two crossed edges $(a,b)$ and $(c,d)$ as shown in Figure~\ref{fig:cosine2}, the cosine force on vertex $a$ is:

\begin{equation}
(-k_{cos}\cos\theta\frac{y_{b}-y_{a}}{d_{ab}}, k_{cos}\cos\theta\frac{x_{b}-x_{a}}{d_{ab}})
\end{equation}

\begin{figure}[h]
\centering
\includegraphics[width=2.2in]{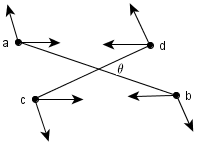}
\caption{The attractive and repulsive cosine forces} \label{fig:cosine3}
\end{figure}

In the third case, the cosine force is divided in two directions. In one direction, the force is applied in an attractive manner and in the other in a repulsive manner. As shown in Figure~\ref{fig:cosine3}, the attractive component of the cosine force on vertex $a$ is:

\begin{equation}
 (k_{cos}\cos\theta\frac{x_{d}-x_{a}}{d_{ad}}, k_{cos}\cos\theta\frac{y_{d}-y_{a}}{d_{ad}})
 \end{equation}
 and the repulsive component can be denoted as:
 \begin{equation}
 (k_{cos}\cos\theta\frac{x_{a}-x_{c}}{d_{ac}}, k_{cos}\cos\theta\frac{y_{a}-y_{c}}{d_{ac}})
 \end{equation}

Our tests showed that all three produced drawings with larger crossing angles than the classical spring algorithm. However, the best overall result was consistently achieved by the parallel cosine force. For the remainder of this paper, we restrict our attention to the parallel cosine force.

It should be noted that in our implementation, edges that intersect at an endpoint are not considered as crossing.

\subsection{The Algorithm}

Given a graph $G=(V,E)$, let $V' \subseteq V$ be the set of vertices in which each vertex has at least one incident edge crossing another; the combined force applied on vertex $v$ is:

\begin{equation}
F(v) = \sum_{(u,v)\in E} f_{s,uv} + \sum_{(u,v)\in V\times V} f_{r,uv} + \sum_{(c,v)\in C\times V'} f_{cos,cv}
\end{equation}
where $f_{s,uv}$ denotes the spring force, $f_{r,uv}$ denotes the repulsive force, $f_{cos,cv}$ denotes the cosine force, and $C \subseteq E\times E$ denotes the set of pairs of edges that cross, in which each pair includes an incident edge of vertex $v$.

BIGCROSS employs the simple \lq\lq follow your nose\rq\rq~approach by iteratively calculating forces and updating the positions of vertices accordingly.

\section{Experimental Results} \label{sec:implement}

Traditionally, in evaluating force-directed methods, the main focus has been on their performance in drawing highly structured graphs, such as planar meshes, trees, hypercubes. However, their usage has gone far beyond those well structured graphs. Nowadays force-directed algorithms are widely used in various application domains to explore real world problems, in which graphs are rarely highly structured. Therefore it is also necessary to investigate how a force-directed algorithm performs with less-structured graphs.

Furthermore, one of known problems with force-directed algorithms is that their performance is not consistent (see, for example, \cite{tunkelang}). The quality of the output heavily relies on the combination of input parameters including initial layout and the choice of constants. Fruchterman and Reingold~\cite{fr} comment that an \lq\lq algorithm should work reasonably well almost always, without the user having to fiddle with options.\rq\rq\,\, Thus we performed experiments based on a set of pre-specified parameters and report our results based on statistical bases.

\subsection{Test Data}

We have tested on five types of sparse ($\vert E\vert \leq 3\vert V \vert$) connected graphs listed below. These graphs were either randomly generated based on well accepted models, taken from benchmark data sets, or from those used in previous papers. A more detailed description is as below.

\textbf{FR graphs}: We collected $38$ graphs from published papers on force directed methods, mostly taken from Fruchterman and Reingold~\cite{fr}. These graphs were all structured. Three different initial placements for each graph were produced for testing. In other words, our test for FR graphs was based on $114$ cases.

\textbf{Planar graphs}: The planar graphs were taken from the RND/BUP set of GDT testsuites~\cite{gdt}. This graph set originally contains $200$ randomly generated undirected planar graphs with size ranging from $10$ to $100$. For our purpose, only graphs with no more than $50$ vertices were used.

\textbf{Random graphs}: We tested on $3$ kinds of random graphs~\cite{hayes,le} generated based on the following three models:
\begin{itemize}
\item Erdos-Renyi model~\cite{erdos}
\item Watts-Strogatz model~\cite{watts}
\item Eppstein-Wang model~\cite{eppstein}
\end{itemize}
For each model, there were $500$ graphs with size ranging from $10$ to $50$.

\subsection{Design}

During the testing, the initial layout for each graph was randomly produced confined within a unit square. Then based on the same layout, we ran BIGCROSS and the classical spring algorithm mentioned in Section~\ref{class} separately.

We ran pilot studies to determined values for the constants: $k_{a}$, $k_{r}$, and $l$ were set to 1, and $k_{cos}$ was set to $1$. These values work reasonably well together most of the time when an initial placement is confined in a unit square.

Each algorithm stopped once the graph system reached a relatively stable status or iterations reached a pre-specified number. The threshold for the stable status was defined as the time when the maximum movement among all vertices in both $x$ and $y$ direction was no more than $0.0005$. The maximum number of iterations was set to $80000$. One exception was that for FR graphs, the threshold for maximum movement was set being $0.00001$ and the maximum number of iterations was $100000$, for best possible final layouts.

Experiments were performed on a 2.4GHz laptop with 2.99GB RAM. The running time and the number of iterations were recorded. The initial and final placements for each graph for both algorithms were measured. The aesthetic criteria we used for graph measurement include:

\begin{enumerate}

\item Number of crossings
\item Average size of crossing angles
\item Standard deviation of crossing angle
\item Average edge length
\item Standard deviation of edge length
\item Angular resolution (the smallest angle size between two edges incident to the same vertex)

\end{enumerate}
Statistical analysis of the results according to these measures is described in the following section.


\begin{table*}[t]
\begin{center}
\caption{Medians of testing measures}
\small

\begin{tabular}{l  c c c c c c c c c c c c c c c c c}

\hline

 &   &\multicolumn{2}{c}{Number of crossings}&  &\multicolumn{2}{c}{Angle size \scriptsize (deg.)}& &\multicolumn{2}{c}{Angle deviation \scriptsize (deg.)} && \multicolumn{2}{c}{Angular resolution \scriptsize (deg.)}\\
\cline{3-4}  \cline{6-7} \cline{9-10} \cline{12-13}
Graph type &  & \it{BIGCROSS} & \it{Classical} & & \it{BIGCROSS} & \it{Classical} & & \it{BIGCROSS} & \it{Classical} & &    \it{BIGCROSS} &    \it{Classical}  \\
\hline

FR               &  & 4   & 4   &    & 76.40 & 72.00 &   & 7.53 & 8.75     && 28.01 & 31.09  \\
Planar           &  & 12  & 12  &    & 72.55 & 68.27 &   & 11.64 & 14.63   && 8.52  & 9.18   \\
Erdos-Renyi      &  & 130 & 135 &    & 66.72 & 62.52 &   & 17.00 & 18.67   && 0.79  & 0.94   \\
Watts-Strogatz   &  & 13  & 12  &    & 73.02 & 67.34 &   & 10.81 & 14.53   && 8.42  & 8.85   \\
Eppstein-Wang    &  & 218 & 247 &    & 65.27 & 62.06 &   & 17.69 & 18.85   && 0.43  & 0.49   \\

\hline

\end{tabular}
\normalsize

\label{tbl:median1}

\end{center}
\end{table*}



\begin{table*}
\begin{center}
\caption{Medians of testing measures}
\small

\begin{tabular}{l  c c c c c c c c c c c c c c c c c}

\hline

 &   &\multicolumn{2}{c}{Edge length}&  &\multicolumn{2}{c}{Edge deviation}& &\multicolumn{2}{c}{Iterations} & &\multicolumn{2}{c}{Running time \scriptsize (sec.)} \\
\cline{3-4}  \cline{6-7} \cline{9-10} \cline{12-13}
Graph type &  & \it{BIGCROSS} & \it{Classical} & & \it{BIGCROSS} & \it{Classical} & & \it{BIGCROSS} & \it{Classical} & &    \it{BIGCROSS} &    \it{Classical}\\
\hline

FR              &   & 1.70 & 1.71   &   & 0.34 & 0.36 &    & 17596& 17056  && 2.18 & 0.25   & \\
Planar          &   & 2.02 & 2.05   &   & 0.48 & 0.51 &    & 6635 & 7175   && 2.69 & 0.73   &  \\
Erdos-Renyi     &   & 1.65 & 1.97   &   & 0.53 & 0.63 &    & 4486 & 5088   && 4.87 & 0.53   &  \\
Watts-Strogatz  &   & 1.87 & 1.92   &   & 0.44 & 0.47 &    & 5887 & 6521   && 2.57  & 0.64  &  \\
Eppstein-Wang   &   & 1.43 & 1.92   &   & 0.53 & 0.64 &    & 4056 & 4712   && 6.61 & 0.50   &  \\

\hline

\end{tabular}
\normalsize

\label{tbl:median2}

\end{center}
\end{table*}


\subsection{Results}

The results are summarized in Tables~\ref{tbl:median1} and~\ref{tbl:median2}. It should be noted that most of the data were not normally distributed, due to the nature of the measurements. Thus the median value, rather than mean, was computed across graphs for accuracy. Accordingly, a nonparametric method called the \emph{Wilcoxcon signed-ranks}\cite{wil} test on paired data was used for statistical analysis.

As can be seen from Tables~\ref{tbl:median1} and~\ref{tbl:median2}, for FR graphs, on average in terms of median, BIGCROSS increased crossing angle by $4.40$ degrees, compared to the classical spring algorithm. It also reduced angle deviation by $1.22$ degree, average edge length by $0.01$ and edge deviation by $0.02$. However, the number of crossings was the same in the two conditions, and angular resolution was decreased by $3.08$ degrees. Wilcoxcon signed-ranks tests indicated that the differences in average angle size, average edge length and edge deviation were statistically significant with $p<0.01$.

For planar graphs, BIGCROSS increased crossing angle by $4.28$ degrees. It also reduced angle deviation by $2.99$ degrees, average edge length by $0.03$ and edge deviation by $0.03$. However, angular resolution was decreased by $0.66$ degree, and the number of crossings was not changed. Wilcoxcon signed-ranks tests indicated that all these differences were statistically significant with $p<0.01$, except that for angular resolution.

For Erdos-Renyi graphs, BIGCROSS increased average crossing angle by $4.20$ degrees. It also reduced angle deviation by $1.67$ degrees and the number of crossings by $5$. Average edge length was decreased by $0.32$, and edge deviation was decreased by $0.08$. However, angular resolution was decreased by $0.15$ degree. Wilcoxcon signed-ranks tests revealed that all these differences were statistically significant with $p\leq 0.001$, except that for angular resolution.

For Watts-Strogatz graphs, BIGCROSS increased average crossing angle by $5.68$ degrees. It also reduced angle deviation by $3.72$ degrees. Average edge length was decreased by $0.05$, and edge deviation was decreased by $0.03$. However, angular resolution was decreased by $0.43$ degree and the number of crossings was increased by 1. Wilcoxcon signed-ranks tests revealed that all these differences were statistically significant with $p\leq 0.001$, except that for angular resolution and for the number of crossings.

For Eppstein-Wang graphs, BIGCROSS increased average crossing angle by $3.21$ degrees. It also reduced angle deviation by $1.16$ and the number of crossings by $29$. Average edge length was decreased by $0.49$, and edge deviation was decreased by $0.11$. However, angular resolution was decreased by $0.06$ degree. Wilcoxcon signed-ranks tests revealed that all these differences were statistically significant with $p \leq 0.01$, except that for angular resolution.

Our emphasis in this study was on the quality of outputs rather than the runtime. Our implementation uses relatively naive methods to find edge crossings and decrease energy. Thus it is expensive in terms of running time in comparison to more refined force-directed methods (for example,~\cite{frick}). Further, the need to compute crossings makes it more expensive than a classical spring method. There are many ways in which the cosine force can be computed more efficiently (note that it is a local force), and we leave this for future study.

\section{Examples}

\begin{figure*}

  \centering
  \subfigure[Initial layout]{
     \label{}
     \includegraphics[height=1.8in]{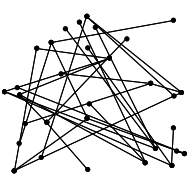}}
  \hspace{0.1in}
  \subfigure[Classical: (1, 82.90, 0, 2.08, 0.47)]{
     \label{}
     \includegraphics[height=1.8in]{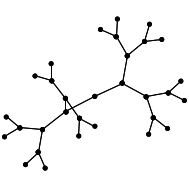}}
  \hspace{0.1in}
  \subfigure[BIGCROSS: (0, 0, 0, 2.07, 0.40)]{
     \label{}
     \includegraphics[height=1.8in]{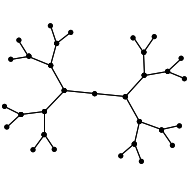}}
  \caption{Tree (graph in Figure 82 from Fruchterman and Reingold~\cite{fr})}
  \label{fig:sym:2}
\end{figure*}

\begin{figure*}

  \centering
  \subfigure[Initial layout]{
     \label{}
     \includegraphics[height=1.8in]{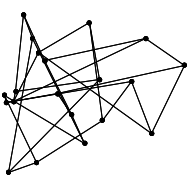}}
  \hspace{0.1in}
  \subfigure[Classical: (7, 81.36, 6.32, 1.99, 0.32)]{
     \label{}
     \includegraphics[height=1.8in]{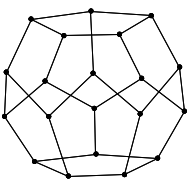}}
  \hspace{0.1in}
  \subfigure[BIGCROSS: (6, 86.62, 3.30, 1.99, 0.30)]{
     \label{}
     \includegraphics[height=1.8in]{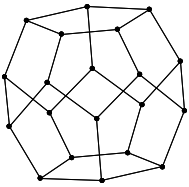}}
  \caption{Dodecahedron (graph in Figure 61 from Fruchterman and Reingold~\cite{fr})}
  \label{fig:sys:10}
\end{figure*}

\begin{figure*}

  \centering
  \subfigure[Initial layout]{
     \label{}
     \includegraphics[height=1.8in]{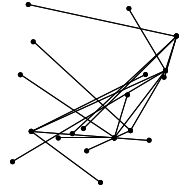}}
  \hspace{0.1in}
  \subfigure[Classical: (0, 0, 0, 2.05, 0.47)]{
     \label{}
     \includegraphics[height=1.8in]{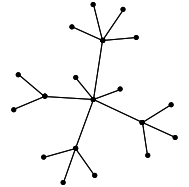}}
  \hspace{0.1in}
  \subfigure[BIGCROSS: (0, 0, 0, 2.05, 0.46)]{
     \label{}
     \includegraphics[height=1.8in]{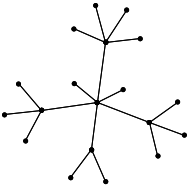}}
  \caption{Tree (graph in Figure 43 from Fruchterman and Reingold~\cite{fr})}
  \label{fig:sys:16}
\end{figure*}

\begin{figure*}

  \centering
  \subfigure[Initial layout]{
     \label{}
     \includegraphics[height=1.8in]{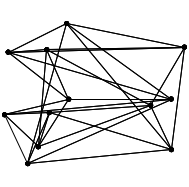}}
  \hspace{0.1in}
  \subfigure[Classical: (20, 60.22, 16.34, 1.64, 0.41)]{
     \label{}
     \includegraphics[height=1.8in]{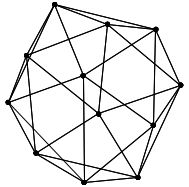}}
  \hspace{0.1in}
  \subfigure[BIGCROSS: (12, 64.49, 4.80, 1.60, 0.29)]{
     \label{}
     \includegraphics[height=1.8in]{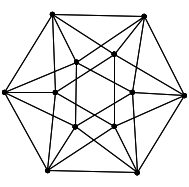}}
  \caption{Icosahedron (graph in Figure 29 from Fruchterman and Reingold~\cite{fr})}
  \label{fig:sys:19}
\end{figure*}

\begin{figure*}

  \centering
  \subfigure[Initial layout]{
     \label{}
     \includegraphics[height=1.8in]{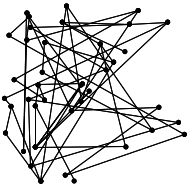}}
  \hspace{0.1in}
  \subfigure[Classical: (10, 64.36, 11.34, 2.07, 0.45)]{
     \label{}
     \includegraphics[height=1.8in]{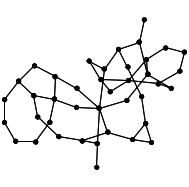}}
  \hspace{0.1in}
  \subfigure[BIGCROSS: (4, 78.36, 2.16, 2.03, 0.34)]{
     \label{}
     \includegraphics[height=1.8in]{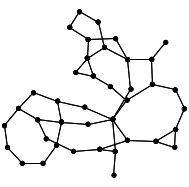}}
  \caption{A planar graph}
  \label{fig:plan:68}
\end{figure*}

\begin{figure*}

  \centering
  \subfigure[Initial layout]{
     \label{}
     \includegraphics[height=1.8in]{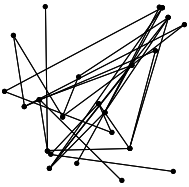}}
  \hspace{0.1in}
  \subfigure[Classical: (8, 62.62, 13.15, 1.96, 0.44)]{
     \label{}
     \includegraphics[height=1.8in]{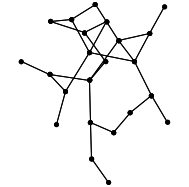}}
  \hspace{0.1in}
  \subfigure[BIGCROSS: (5, 78.06, 4.22, 1.96, 0.43)]{
     \label{}
     \includegraphics[height=1.8in]{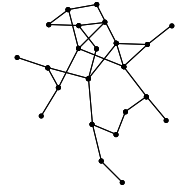}}
  \caption{A graph of Erdos-Renyi model}
  \label{fig:er:445}
\end{figure*}

\begin{figure*}

  \centering
  \subfigure[Initial layout]{
     \label{}
     \includegraphics[height=1.8in]{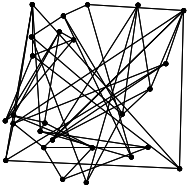}}
  \hspace{0.1in}
  \subfigure[Classical: (15, 55.66, 21.73, 1.79, 0.49)]{
     \label{}
     \includegraphics[height=1.8in]{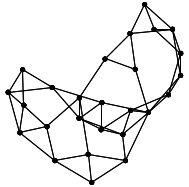}}
  \hspace{0.1in}
  \subfigure[BIGCROSS: (7, 78.89, 7.54, 1.78, 0.41)]{
     \label{}
     \includegraphics[height=1.8in]{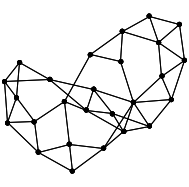}}
  \caption{A graph of Watts-Strogatz model}
  \label{fig:wat:303}
\end{figure*}

\begin{figure*}

  \centering
  \subfigure[Initial layout]{
     \label{}
     \includegraphics[height=1.8in]{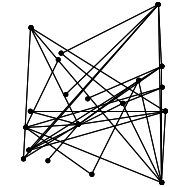}}
  \hspace{0.1in}
  \subfigure[Classical: (26, 62.21, 16.15, 1.83, 0.60)]{
     \label{}
     \includegraphics[height=1.8in]{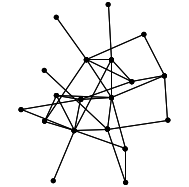}}
  \hspace{0.1in}
  \subfigure[BIGCROSS: (22, 70.57, 15.25, 1.77, 0.51)]{
     \label{}
     \includegraphics[height=1.8in]{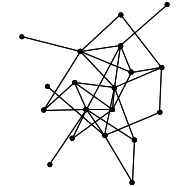}}
  \caption{A graph of Eppstein-Wang model}
  \label{fig:epps:128}
\end{figure*}

\begin{figure*}

  \centering
  \subfigure[Initial layout]{
     \label{}
     \includegraphics[height=1.8in]{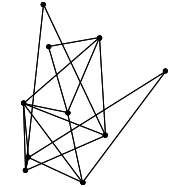}}
  \hspace{0.1in}
  \subfigure[Classical: (0, 0, 0, 1.64, 0.06)]{
     \label{}
     \includegraphics[height=1.8in]{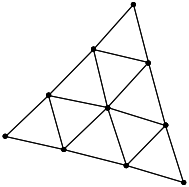}}
  \hspace{0.1in}
  \subfigure[BIGCROSS: (3, 72.69, 0.74, 1.61, 0.28)]{
     \label{}
     \includegraphics[height=1.8in]{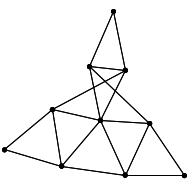}}
  \caption{A triangulated triangle (graph in Figure 65 from Fruchterman and Reingold~\cite{fr})}
  \label{fig:sys:6}
\end{figure*}

From section~\ref{sec:implement}, it is clear that BIGCROSS outperforms the classical spring algorithm. In this section, examples are presented in Figures~\ref{fig:sym:2} -~\ref{fig:sys:6}, all of which are taken from our experiments. Coupling with each resulting drawing, measurement data are provided in the format of (the number of crossings, average crossing angle, angle deviation, average edge length, edge deviation). Where the number of crossings is $0$, both average crossing angle and angle deviation are set to $0$.

It is important to note that our conclusion that BIGCROSS performs better is based on statistical analysis. Given a set of pre-specified constant values, there are cases in which the classical spring algorithm performs better, as shown in Figure~\ref{fig:sys:6}.

\section{Conclusion and Future Work} \label{sec:clu}

We have introduced and implemented the cosine force in a force-directed algorithm to increase crossing angles. The experimental results demonstrate that applying cosine force in addition to traditional spring forces in graph drawing not only increases crossing angles, but also surprisingly yields overall more aesthetically pleasing drawings. Our main contributions are:
 \begin{enumerate}
\item Introduction of the cosine force.
\item Implementation of a new aesthetic criterion - crossing angle - in an algorithm.
\item Introduction of a new force-directed method that produces drawings with some important aesthetics being significantly improved, in comparison to the classical spring algorithm.
\item Statistical evaluation of the algorithm against different types of graphs.
\end{enumerate}

For future work, we note that that it is feasible to speed up BIGCROSS using a fast crossing detection method~\cite{balaban} and/or fast energy minimization methods (for example, FADE~\cite{quigley}). Further, the cosine force can be easily integrated into other force-directed methods, such as Hu-Koren methods to reduce warping~\cite{hu}.

Further, our results indicate that a cosine force may be a good way to increase angular resolution at vertices. Two edges incident to a vertex can be considered as being crossed at the vertex. Therefore, applying cosine force on adjacent edges is a natural extension of what has been presented in this paper. Maximizing vertex angular resolution is particularly interesting since crossing angle competes with angular resolution when two adjacent edges cross the same edge.

\end{document}